\UseRawInputEncoding
\documentclass[pra,12pt,tightenlines,nofootinbib]{revtex4} 

\usepackage{amsmath}
\usepackage{graphics, graphicx}
\usepackage{epsfig}
\usepackage{ulem}
\usepackage{color}

\newcommand{\ttle}[1]{{\it #1}}

\begin{document}

\title{What Do We Learn by Deriving Born's Rule?}

\author{James  Hartle}
\affiliation{Department of Physics, University of California, Santa Barbara,  93106, USA}
\bibliographystyle{unsrt}

\begin{abstract}{Derivations of  Born's Rule are of interest because they tell us what's connected to what in quantum mechanics if we ever need to modfify or change it.}
\end{abstract}

\vskip.8in
\centerline{} 
\vspace{1cm}

\maketitle


\section{Introduction}
\label{intro}

Born's rule (BR)   connects the quantum state of a system to the probabilities for the outcomes of measurements carried out on it. It was introduced in a footnote of a paper by Born published in 1926 \cite{Bor26}. The author's historical scholarship is poor, but it seems from that time to the late '60s of the last century no one thought to derive Born's rule from something else. About  that time there appeared three similar efforts to derive it from other assumptions \cite{Fink63,Har68,Gra73}. Since that time there have been numerous other derivations based on various assumptions, principles, and formulations of quantum mechanics. (For the ones known to the author see references 1-20  in the bibliography. Many more can probably be found in the reviews. \cite{Land,vaid}. Further, Born's rule has, in effect, been extended far beyond simply supplying probabilities for measurement outcomes.   For instance it has been extended to the calculation of the probabilities for for a simple quantum (wave packet) model of the  orbit of the Moon when it is not receiving attention from observers, and to quantum models of density fluctuations in the very early Universe when there were no observers around and no measurements being carried out (e.g. \cite{HHH10a}). 

We should stress that by Born's rule we fo not mean merely a connection of probability to the square of an amplitude. We mean that in a laboratory measurement situation.  For issues in extending the idea beyond that see \cite{Page-br}.


There have been many derivations of BornÕs rule. Some of these are  referenced in the bibliography.  Some are restricted to laboratory quantum mechanics some are not. They differ in their level of rigor, the assumptions they make, the mathematics used, the different applications envisaged, etc. Some use symmetry, some calculate frequencies, some count Everett branches, some count experiments, some count multiverses, 
some count observers, some posit non-linear quantum dynamics etc. Some invoke multiverses, some many worlds, some, spatially infinite universes, some count observers, some are based on  GleasonÕs theorem, and some  invoke human decision theory.

All  derivations of Born's Rule require some assumptions. Roughly these can be divided into  assumptions  of symmetries, assumptions about Hilbert space, and assumptions about which alternatives are certain, 

This essay is not an attempt at a critical comparison of all these efforts, not least because the author is not familiar enough with all of them. Rather it addresses the question of what do we learn from all these derivations and why are they of interest now.

It's important to emphasize that a derivation of Born's Rule is not required  to formulate CGM. But derivations have other uses when we go beyond CQM as described in Sections \ref{beyondlab}  and \ref{genqm}

\section{Laboratory (Copenhagen) Quantum Mechanics (CQM) }
\label{labqm}

The overwhelming success  of  CQM  (including Born's Rule)  in measurement situations makes it clear that there is no {\it need} to derive Born's Ru;e  in that domain. It is part of a consistent theoretical structure --- Copenhagen quantum mechanics\footnote{At the present level of discussion we mean the same thing by `laboratory quantum mechanics', `Copenhagen quantum mechanics', `textbook quantum mechanics', `usual quantum mechanics'  and `the approximate quantum mechanics of measured subsytems'. Consistency is the ....} --- whose predictions agree accurately with delicate experiments. So what  do we learn from a derivation of Born's Rule  from other assumptions?

One thing that we learn is what's connected to what in quantum mechanics. Every derivation makes some set of assumptions. For instance, often some kind of symmetry is assumed between  physical situations which whose probabilities must be equal if probabilities there are. If there are just two situations the probability for each would be $1/2$. Further arguments bootstrap this equivalence to give probabilities for inequivalent physical situations.
This kind of derivation connects Born's Rule  to a notion of physical equivalence. 

The variety of assumptions that have been used to derive Born's Rule  in this way show that  many parts of laboratory quantum mechanics are connected. If one part is changed the other must also change. From this point of view there is no point in debating the relative merits of the assumptions made in different derivations. Within the context of laboratory quantum mechanics such debates are about personal prior probabilities. And, like most debates on personal priors, they are unlikely to be settled. 

Of course, all the derivations of Born's Rule rely on assumptions.  Roughly these divide into a few categories listed below with a representative paper for each:
\begin{itemize}
\item{Calculating the frequencies of occurance in  ensembles of identical situations e.g.  \cite{Har68}}.
 \item{Counting Everett branches of  state of the Universe,, e.g. }\cite{saund}
 \item{Deduction from the structure of Hilbert space,  e.g } \cite{Harup} 
 \item{Deduction from an assumption of what's certain, e.g.} \cite{Deu99}
\end{itemize} 

It's important to emphasize that a derivation of Born's Rule  is not required to formulate CQM as we understand it today.  What then are the motivations that various authors find to  provide them?  Improving pedagogy might be one reason.  Support for positing higher prior probabilities  for a theory with fewer assumptions might be another. But perhaps the main reason is this::  We understand a thing better when we understand the connections between its pieces and can describe  those connections in different ways.

However  derivations of Born's rule become important if experiment  and/or observation forces us to modify  CQM or generalize it to physical realms beyond measurement situations as  is described in the next section.

Were  only connections within  CQM  of interest they could be derived in another way.
Assume Born's rule and see if it implies the connections. In the next Section we will see why derivations are of interest in generalizing quantum theory to apply beyond the laboratory.

\section{Beyond CQM}
\label{beyondlab} 
The Copenhagen (textbook) quantum mechanics of laboratory measurements carried out by observers  CQM is arguably the most successful predictive framework in physics in the last four-hundred years.  But it  must be generalized to apply to quantum systems that are not concerned exclusively with laboratory measurements.  In particular it must be generalized to closed systems  like the Universe as a whole where observers and their apparatus are inside the closed system and not somehow outside looking at it. 
 But there are at least two reasons why it necessary to go beyond  CQM that are described in the next two sections.

\subsection{Derivations of BR Limit Testing Laboratory Quantum Mechanics}
\label{testing}
To test  CQM it would be very useful for motivating and analyzing experiment to go beyond CQM to find theories that were parametrically close to quantum mechanics but not quantum mechanics itself. The connections revealed by derivations of Born's Rule make this difficult to to do.   For example we could not just replace Born's rule with one in which the norm of the state was raised to the $2+\epsilon$ power with small $\epsilon$  and test for the value 
$\epsilon$. It would be inconsistent with the connections,   As Steve Weinberg says ``It is striking that  it has so far not been possible to find a logically consistent theory that is close to quantum mechanics  other than quantum mechanics itself.''\cite{dreams}.

\section{Generalizing CQM}
\label{genqm} 

 CQM  must be generalized to apply to quantum systems that are not concerned exclusively with laboratory measurements.  In particular it must be generalized to closed systems  like the Universe as a whole where observers and their apparatus are inside the closed system and not somehow outside looking at it. 

 At  a fundamental level, the Universe is quantum mechanical system with a quantum state $\Psi$. A theory of such a closed system  consists of a theory of quantum dynamics $I$,a  theory of the state $\Psi$ like the no-boundary wave function \cite{NBWF,WINBWF} , and a quantum framework $\cal Q$ for calculating the probabilities of the the individual members of decoherent sets of suitably coarse-grained histories  the closed systems histories describing the  what goes on inside  the closed system.   The theory is $(I,\Psi,{\cal Q})$.
 
 We test and utilize this theory by predicting probabilities for what we will observe in the universe and comparing them with actual observations.
The classical behavior of the geometry and fields in a quantum universe is an approximation in this quantum framework that is appropriate only when the probabilities are high for appropriately coarse grained histories that exhibit correlations in time governed by deterministic classical laws (e.g. [24]).
The quantum mechanics of laboratory measurements described in Section \ref{labqm} is a feature of  the Universe  that emerges from ( $I$,$\Psi$,$\cal Q$) somewhat after the big bang along with classical spacetime, isolated systems, fitness landscapes, biological evolution, information gathering and utilizing systems (IGUSes) like ourselves. etc,etc
\cite{HarWUComp}

\section{Decoherent Histories  Quantum Mechanics}
\label{BRfrmDH}
We now have a formulation of  quantum mechanics for closed systems like the Universe called decoherent (or consistent) histories quantum mechanics.(DH)  It is the work of many  e.g \cite{classicDH}.  For  reviews see \cite{Harhouch,Har07}. A tutorial can be found in \cite{HarQMCS}, For the author's original contribution with Murray Gell-Mann see \cite{GH90}.. 

From $(I,\Psi, {\cal Q} )$  DH predicts probabilities for the individual members of decoherent sets of alternative coarse-grained time histories of the closed system. Thus, DH enables quantum cosmology.

Observers and their measuring apparatus are physical systems within the Universe. By studying their interactions in a closed system with a flat classical spacetime  e.g
\cite{Roig} CQM can be understood as an approximation to DH appropriate for measurement siguations.  It is  possible to start even farther back with the quantum probabilities for the classical spacetime assumed in formulating CQM with dynamics in $I$ that includes quantum spacetime (i.e. quantum gravity) , see e.g. \cite{Harhouch} .

Thus CQM can be  seen as an approximation to DH.  And thus, we have yet another derivation of Born's rule. --- not from assuming just CQM but rather  starting from the more general DH and deriving CQM.

\section{Connections}
\label{connections}
An important result of the various derivations of BR in the context CQM  is that we  learn how some things are connected to others in CQM.  There is something similar in DH.  DH itself can be formulated in  different ways. For example, DH  can be formulated using extended probabilities that are sometimes negative \cite{Harextp} ,using records of what happened \cite{HarRecords},, and using the linear positivity decoherence condition \cite{HarVirp}. There are many other examples.

\section{Conclusion} By deriving Born's Rule in  many ways we learn much about the structure of its familiar formulation for measurement situations (CQM).  But beyond that, we learn much about  constraints on alternatives to CQM  useful in testing quantum  theory and on its generalizations to  other realms of physics such as the cosmology of our Universe  
\cite{HarJer, Harhouch} .

\noindent{\bf Acknowledgments:} 

The work of JH was supported in part by an US NSF grant PHY-18-8018105.
The author has benefited over the years with many discussions on these questions with Mark Srednicki and the late Murray Gell-Mann. 

This paper is dedicated to Wojtek Zurek for his 70th birthday,  for the many pr supportive discussions we have had, but more generally for the many contributions he has made to our understanding of quantum theory  not least his own derivation of Born's Rule  \cite{Zur05}. 

\eject

\end{document}